\documentclass[twocolumn,showpacs,superscriptaddress,preprintnumbers,floatfix,amsmath,amssymb,prb]{revtex4}

\usepackage{graphicx}
\usepackage{dcolumn}
\usepackage{bm}
\usepackage{amsfonts}
\usepackage[below]{placeins}
\usepackage{subfigure}

\newcommand{\bfpsi}{\mbox{\boldmath$\psi$}}

\begin{document}


\title{Complex magnetic behavior of the sawtooth Fe chains in Rb$_{2}$Fe$_{2}$O(AsO$_{4}$)$_{2}$}

\author{V. Ovidiu Garlea}
 \email{garleao@ornl.gov}
\affiliation{Quantum Condensed Matter Division, Oak Ridge National Laboratory, Oak
Ridge, TN 37831, USA}
\author{Liurukara D. Sanjeewa}
\affiliation{Department of Chemistry, Clemson University, Clemson, South Carolina 29634-0973, USA}
\author{Michael A. McGuire}
\affiliation{Materials Science and Technology Division, Oak Ridge National Laboratory, Oak Ridge, Tennessee 37831,
USA}
\author{Pramod Kumar}
\affiliation{Indian Institute of Information Technology, Allahabad U.P. 211012, India}
\author{Dino Sulejmanovic}
\affiliation{Department of Chemistry, Clemson University, Clemson, South Carolina 29634-0973, USA}
\author{Jian He}
\affiliation{Department of Physics and Astronomy, Clemson University, Clemson, South Carolina 29634-0978,USA}
\author{Shiou-Jyh Hwu}
\affiliation{Department of Chemistry, Clemson University, Clemson, South Carolina 29634-0973, USA}

\date{\today}

\begin{abstract}
Results of magnetic field and temperature dependent neutron diffraction and magnetization measurements on oxy-arsenate Rb$_{2}$Fe$_{2}$O(AsO$_{4}$)$_{2}$ are reported. The crystal structure of this compound contains pseudo-one-dimensional [Fe$_{2}$O$_{6}$]$^\infty$ sawtooth-like chains, formed by corner sharing isosceles triangles of $Fe^{3+}$ ions occupying two nonequivalent crystallographic sites. The chains extend infinitely along the crystallographic $b$-axis and are structurally confined from one another via diamagnetic (AsO$_{4}$)$^{3-}$ units along the $a$-axis, and Rb$^+$ cations along the $c$-axis direction. Neutron diffraction measurements indicate the onset of a long range antiferromagnetic order below approximately 25 K. The magnetic structure consists of ferrimagnetic chains which are antiferromagnetically coupled with each other. Within each chain, one of the two Fe sites carries a moment which lies along the \emph{b}-axis, while the second site bears a canted moment in the opposite direction. Externally applied magnetic field induces a transition to a ferrimagnetic state, in which the coupling between the sawtooth chains becomes ferromagnetic. Magnetization measurements performed on optically-aligned single crystals reveal evidence for an uncompensated magnetization at low magnetic fields that could emerge from to a phase-segregated state with ferrimagnetic inclusions or from antiferromagnetic domain walls. The observed magnetic states and the competition between them is expected to arise from strongly frustrated interactions within the sawtooth chains and relatively weak coupling between them.
\end{abstract}

\pacs{75.50.Gg, 75.60.Jk, 75.25.-j, 71.27.+a, 61.05.F-}

\maketitle

\section{Introduction}

Low-dimensional magnetic materials have drawn continued attention in condensed matter physics, owing to their
distinct electronic and magnetic properties. In particular, the oxyanion systems based on transition metal
($M$) oxides sublattices that are magnetically isolated by closed shell nonmagnetic oxyanions (SiO$_4^{4-}$,
PO$_4^{3-}$, AsO$_4^{3-}$) show great potential for exploring and characterizing  new emergent phenomena.\cite{Hwu} This class of compounds has been shown to exhibit a variety of magnetic ground states ranging from  ferromagnetism, to ferrimagnetism and antiferromagnetism.\cite{magnetism} On the other hand, due to their impressive structural diversity in terms of architecture these materials have shown potential applications in the fields of energy storage, chemical sensors, catalysis and biomedical research.\cite{Wurm,Uebou,Masquelier,Padhi}

\begin{figure}[tbp]
\includegraphics[width=3.4in]{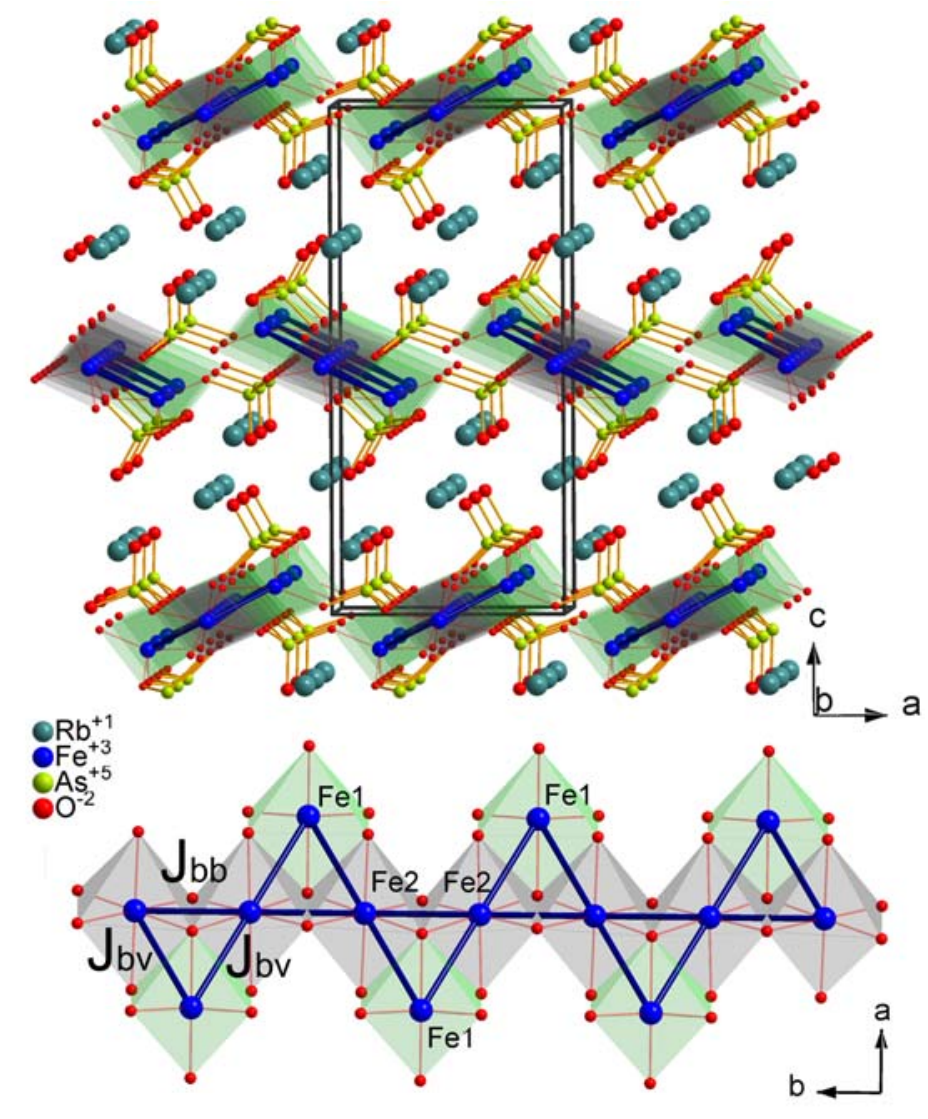}
\caption{\label{structure}(Color online) (top) Polyhedral view of the extended structure of
Rb$_{2}$Fe$_{2}$O(AsO$_{4}$)$_{2}$ showing the structural confinement of the chains. (bottom) Partial structure of the [Fe$_{4}$O$_{14}$]$^\infty$ sawtooth chain defined by unequal base-base ($J_{bb}$) and base-vertex ($J_{bv}$) exchange interactions.}
\end{figure}

Recent efforts have been directed towards the synthesis of new members of the family of oxy-arsenates with
general formula $A_{2}M_{2}$O(AsO$_{4}$)$_{2}$, where \emph{A} = K, Rb, and \emph{M} is a transition metal. These compounds exhibit intriguing one-dimensional triangular-shaped chains made of edge-sharing \emph{MO}$_6$
octahedra (see Fig.~\ref{structure}). Antiferromagnetic sawtooth chain model (or $\Delta$-chain), as that realized in these systems, is known to be highly frustrated and has drawn considerable attention from a theoretical standpoint.\cite{Sen, Nakamura, Blundell} For example, the ground state of the \emph{S}=1/2 sawtooth chain is two-fold degenerate with either left pair or right pair of spin at each triangle forming spin-singlet states. The lowest excitation in a chain with periodic boundary conditions is given by a \emph{kink-antikink} pair which has a dispersionless gap $\Delta$E $\simeq$0.234$J$ where $J$ is the coupling between pairs of spins.\cite{Sen, Nakamura} Experimental realization of magnetic lattices corresponding to the sawtooth chain is only found in a very small number of compounds, including the delafossite YCuO$_{2.5}$,\cite{Garlea} euchroite Cu$_{2}$(AsO$_4$)(OH)$\cdot$3H$_2$O~\cite{euchroite} and Zn$L_2$S$_4$ olivines, with $L$=Er, Tm, Yb.~\cite{Lau}

We report here on the magnetic behaviour of oxy-arsenite Rb$_{2}$Fe$_{2}$O(AsO$_{4}$)$_{2}$, containing sawtooth chains of  magnetic Fe$^{3+}$ ions in high-spin state $S$=5/2. Although this compound was first reported five years ago,~\cite{Chang} a thorough investigation of its magnetic properties has not been conducted so far. The development of well-controlled methods for the growth of high quality single crystals has allowed us for detailed orientation-dependent magnetic measurements. Magnetization measurements have been complemented with x-ray and neutron diffraction measurements. Temperature and magnetic field dependent neutron diffraction measurements demonstrate conclusively the development of a non-collinear antiferromagnetic ground state below approximately 25 K, which under relatively low magnetic fields, transforms to a ferrimagnetic state. These transitions are also observed in \emph{dc} magentization measurements, which reveal strong magnetic anisotropy and complex field and temperature dependent behaviors associated with the multiple magnetic states.

\section{Experimental details}

Single crystals of Rb$_{2}$Fe$_{2}$O(AsO$_{4}$)$_{2}$ were grown using flux methods in the RbCl/RbI (50:50 wt\%)
molten-salt media. The reactants KO$_{2}$ (Alfa Aesar, 96.5\%), Fe$_{2}$O$_{3}$ (Alfa Aesar, 99.945\%) As$_{2}$O$_{5}$ (Alfa Aesar, 99.9\%) were mixed and ground together in a nitrogen-purged dry box. The reaction mixture was sealed in an evacuated fused-silica ampoule and then heated to 650 $^\circ$C for one day, followed by another heating to 800 $^\circ$C at 1 $^\circ$C/min for two days, slowly cooled to 300 $^\circ$C at 0.05 $^\circ$C/min, and then furnace-cooled to room temperature. Brown columnar crystals (up to about 2.6 mm x 0.6 mm x 0.4 mm) were recovered by washing the product with de-ionized water using vacuum filtration method.
Single crystals were physically examined and selected under an optical microscope equipped with a polarizing light attachment. The crystals were then mounted on a glass fiber and single crystal x-ray diffraction data was collected using a Rigaku Mercury CCD detector on an AFC-8S diffractometer equipped with graphite monochromated Mo K$\alpha$ radiation ($\lambda$ = 0.7107~{\AA}). The structure was solved by direct methods using the SHELX-97\cite{shelex} program and refined on F$^2$ by least-squares, full-matrix technique.

Polycrystalline samples were prepared by the solid state reaction of stoichiometric amounts of Rb$_{2}$CO$_{3}$
(Alfa Aesar, 99.8\%), Fe$_{2}$O$_{3}$ (Alfa Aesar, 99.945\%) and (NH$_{4}$)H$_{2}$AsO$_{4}$ (Alfa Aesar, 99.9\%). The mixture was loaded into an alumina crucible, heated to 800 $^\circ$C at a rate of 2 $^\circ$C/min, and held at that temperature for 2 days before being furnace cooled to room temperature. All polycrystalline samples were characterized using Rigaku Ultima IV multipurpose X-ray diffraction system.

Temperature and field-dependent magnetic measurements were carried out with a Quantum Design Magnetic Property Measurement System (MPMS). The measurements were taken on collections optically co-aligned single crystals at temperatures from 2 K to 300 K in the applied fields up to 50 kOe. Additional magnetization measurements were performed on polycrystalline sample using a physical property measurement system (PPMS) at fields up to 120 kOe.

Neutron diffraction measurements were carried out using the HB2A high-resolution powder diffractometer at the High Flux Isotope Reactor (HFIR). Diffraction patterns were collected using the 2.408~{\AA} wavelength produced by the (113) reflection of a vertically focusing germanium wafer-stack monochromator. The beam collimation was
12$^{\prime}$--40$^{\prime}$--6$^{\prime}$. More details about the HB2-A instrument and data collection strategies can be found in Ref.~\onlinecite{HB2A}. For the measurements, polycrystalline samples with a total mass of approximately 5 grams were compacted in pellets to prevent grain reorientation by the applied field. The pellets were loaded into a cylindrical vanadium can, and placed in a vertical-field cryomagnet. Data were collected at various temperatures from 40 K and down to 4 K, and magnetic fields up to 40~kOe. Rietveld refinements were carried out with the FullProf Suite program.~\cite{fullprof}

\begin{table}[tbp]
\caption{\label{table1} Refined structural parameters of Rb$_{2}$Fe$_{2}$O(AsO$_{4}$)$_{2}$ from single-crystal
x-ray data collected at room temperature.}
\begin{ruledtabular}
\begin{tabular}{cccccc}
\centering{Atom (Wyck.)} & $x$ & $y$ & $z$ & U$_{eq}$ \\[3pt]
\hline
\centering {Rb1~~($4c$)}& 0.8888(1) & 1/4 & 0.8628(1) & 0.023(1) \\[3pt]
\centering {Rb2~~($4c$)}& 0.0379(1) & 1/4 & 0.2679(1) & 0.028(1) \\[3pt]
\centering {Fe1~~($4c$)}& 0.2844(1) & 1/4 & 0.4347(1) & 0.009(1) \\[3pt]
\centering {Fe2~~($4b$)}& 0 & 0 & 1/2 & 0.009(1) \\[3pt]
\centering {As1~~($4c$)}& 0.8204(1) & 1/4 & 0.6426(2) & 0.010(1) \\[3pt]
\centering {As2~~($4c$)}& 0.6758(1) & 1/4 & 0.4440(2) & 0.008(1) \\[3pt]
\centering {O1~~($4c$)}& 1.0580(1) & 1/4 & 0.4375(3) & 0.009(1)  \\[3pt]
\centering {O2~~($4c$)}& 0.9610(1) & 1/4 & 0.5760(3) & 0.010(1) \\[3pt]
\centering {O3~~($8d$)}& 0.7666(1) & 0.0108(1) & 0.4785(2) & 0.015(1) \\[3pt]
\centering {O4~~($4c$)}& 0.4923(1) & 1/4 & 0.4744(4) & 0.014(1) \\[3pt]
\centering {O5~~($8d$)}& 0.7043(1) & 0.0127(1) & 0.6338(2) & 0.018(1) \\[3pt]
\centering {O6~~($4c$)}& 0.6843(1) & 1/4 & 0.3556(3) & 0.018(1) \\[3pt]
\centering {O7~~($4c$)}& 0.9130(2) & 1/4 & 0.7197(3) & 0.025(1) \\[3pt]
\hline \\
\multicolumn{5}{c}\centering{$Pnma$,~$a$ = 8.5331(2)~\AA,~$b$ = 5.7892(2)~\AA,~$c$ = 18.611(4)~\AA}\\
\multicolumn{5}{c}{R$_{1}$ = 0.0347, wR$_{2}$ = 0.0936, GOF = 1.038}\\
\end{tabular}
\end{ruledtabular}
\end{table}

\begin{table}[tbp]
\caption{\label{table2}Selected bond distances and angles in Rb$_{2}$Fe$_{2}$O(AsO$_{4}$)$_{2}$, as determined from
x-ray analysis.}
\begin{ruledtabular}
\begin{tabular}{cccc}
Fe1 - O1 (x 1)& 1.932(6)~\AA& Fe2 - O1 (x 2) & 1.922(4)~\AA  \\[3pt]
Fe1 - O3 (x 2)& 2.254(4)~\AA& Fe2 - O2 (x 2) & 2.051(4)~\AA  \\[3pt]
Fe1 - O4 (x 1)& 1.923(6)~\AA & Fe2 - O3 (x 2) & 2.032(4)~\AA  \\[3pt]
Fe1 - O5 (x 2)& 1.986(4)~\AA & & \\[3pt]
\hline\\
\multicolumn{4}{c}{\emph{d}(Fe1 - Fe2) = 3.076(1)~\AA~ $\Leftrightarrow$ $J_{bv}$  via O1, O3}\\[3pt]
\multicolumn{2}{c}{Fe1 - O3 - Fe2 = 91.5(1)$^{\circ}$} & \multicolumn{2}{c} {Fe1 - O1 - Fe2 =105.9(1)$^{\circ}$}
\\[3pt]
\multicolumn{4}{c}{\emph{d}(Fe2 - Fe2) = 2.895(1)~\AA~ $\Leftrightarrow$ $J_{bb}$ via O1, O2}\\[3pt]
\multicolumn{2}{c}{Fe2 - O1 - Fe2 = 97.7(1)$^{\circ}$} & \multicolumn{2}{c} {Fe2 - O2 - Fe2 = 89.7(1)$^{\circ}$}
\\[3pt]
\end{tabular}
\end{ruledtabular}
\end{table}

\section{Results and Discussion}
\subsection{Crystal structure}

The single crystal X-ray and neutron powder diffraction data were well fit by using an orthorhombic
structure of space group $Pnma$, similar to that proposed by Chang \emph{et al.}.~\cite{Chang} The refined structural parameters of Rb$_{2}$Fe$_{2}$O(AsO$_{4}$)$_{2}$, such as atomic coordinates and displacement parameters are listed in
Table~\ref{table1}, while selected bond lengths and angles are given in Table~\ref{table2}. Rietveld analysis indicate that the powder samples contain a Fe$_2$O$_3$ impurity phase in an amount of less than 3\% weight.

The structure of Rb$_{2}$Fe$_{2}$O(AsO$_{4}$)$_{2}$ consists of iron-arsenate layers parallel to the $ab$ plane,
separated by Rb$^+$ cations. In each layer, there are two crystallographically distinct Fe$^{3+}$ sites (Fe1 and
Fe2) which are bridged together to form triangle-based chains, as shown in Fig.~\ref{structure}. These [Fe$_{4}$O$_{14}$]$^\infty$ chains extend along the crystallographic $b$-axis, and are structurally confined from one another via diamagnetic As$^{5+}$ cations in AsO$_4$ tetrahedral units along the crystallographic $a$-axis (Fig.~\ref{structure}). Due to the non-magnetic nature of As$^{5+}$ cation, the magnetic interactions between adjacent [Fe$_{4}$O$_{14}$]$^\infty$ $\Delta$-chains via Fe-O-As-O-Fe connection are expected to be subdominant compared to the interactions within the chains.

A comparison of the refined bonds and angles in Rb$_{2}$Fe$_{2}$O(AsO$_{4}$)$_{2}$ indicates different
interactions occurring between spins on the base ($J_{bb}$ coupling) and between the base-vertex ($J_{bv}$) of the triangles. The Fe1–-Fe2 and the Fe2-–Fe2 distances are 3.076(1) Å and 2.895(1) Å respectively. The bond angles Fe2--O1--Fe2 and Fe2--O2--Fe2 defining the super-exchange paths along the base of triangles ($J_{bb}$) are 97.7(3)$^{\circ}$ and 89.7(7)$^{\circ}$, respectively, while the base-vertex angles ($J_{bv}$) Fe1-O1-Fe2 and Fe1-O3-Fe2 are 105.9(1)$^{\circ}$ and 91.5(1)$^{\circ}$.

\subsection{Macroscopic magnetic measurements}

\begin{figure}[tbp]
\includegraphics[width=3.5in]{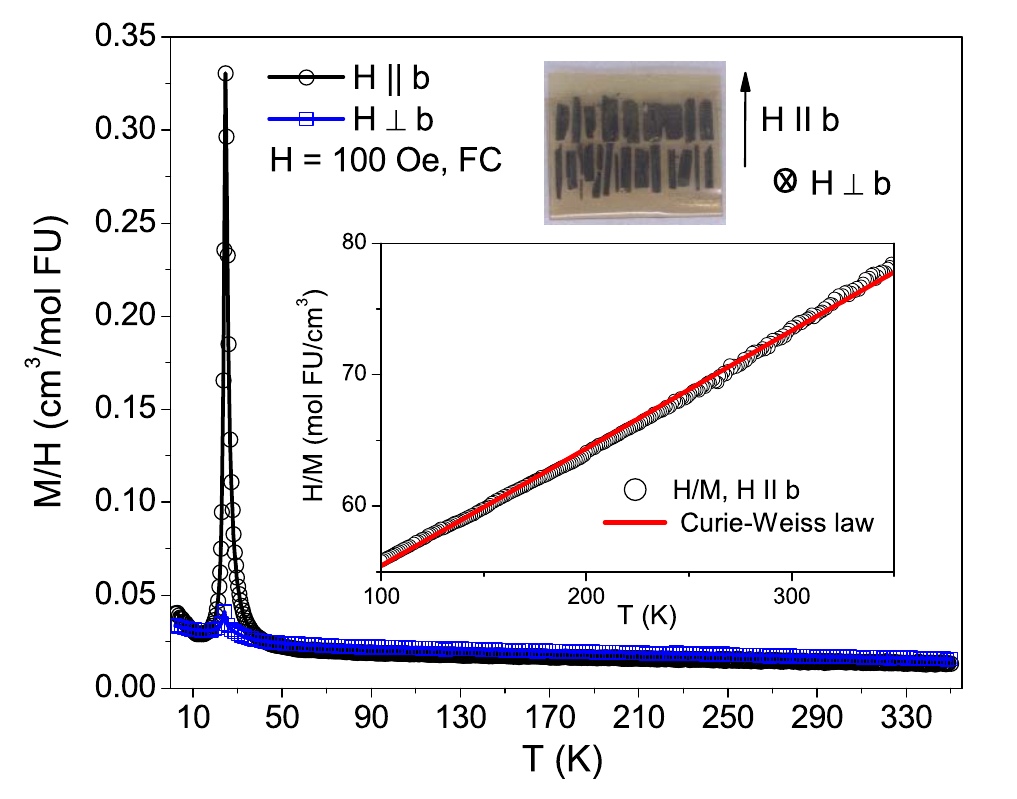}
\caption{\label{suscept} Evolution of the magnetic susceptibility (M/H) as a function of temperature, for H = 100 Oe applied parallel and perpendicular to the $b$-axis. The inserted picture displays the single crystals assembly used in the measurements. The inset shows the inverse susceptibility versus temperature above the transition temperature, and a linear (Curie-Weiss) fit.}
\end{figure}

Magnetization measurements were performed on single crystals with the $b$-axis (chain direction) aligned parallel and perpendicular to the magnetic field. Due to the small size of the crystals, several specimens were optically co-mounted on a Kapton tape as displayed in Fig.~\ref{suscept}. Two samples were prepared with total masses of 4.0 and 8.4 mg. The measured magnetization was confirmed to be consistent between the two samples. The temperature dependence of magnetic susceptibility ($\chi=M/H$) measured with an applied field of 100 Oe, is shown in Fig.~\ref{suscept}. The sharp rise in magnetization data reveals the onset of long range magnetic order below approximately 25 K. There is strong anisotropy between the two crystal orientations near the transition temperature. The inverse susceptibilty (H/M) for temperature well above this temperature is shown in the inset, along with a fit using the Curie Weiss model, M/H = $C$/(T - $\Theta_{CW}$). The data are not strictly linear over this wide temperature range. A better fit can be obtained by including a temperature independent term; however, such a fit returns unreasonable values for the fitting parameters. The observed deviation from Curie Weiss behavior may indicate some short ranged or low dimensional magnetic ordering above the long-ranged ordering transition at 25 K. Using the fit shown in Fig.~\ref{suscept}, an effective moment of 6.7(1)~$\mu_B$ and a Weiss temperature of -510 K are determined. Restricting the fitted range to the more linear region between 250 and 350 K gives an effective moment of 6.3~$\mu_B$ and a Weiss temperature of -426 K. These effective moments are somewhat larger than the spin only value for high spin $d^5$, 5.9~$\mu_B$. The large and negative Weiss temperatures indicates strong antiferromagnetic interactions.

\begin{figure}[tbp!]
\includegraphics[width=3.5in]{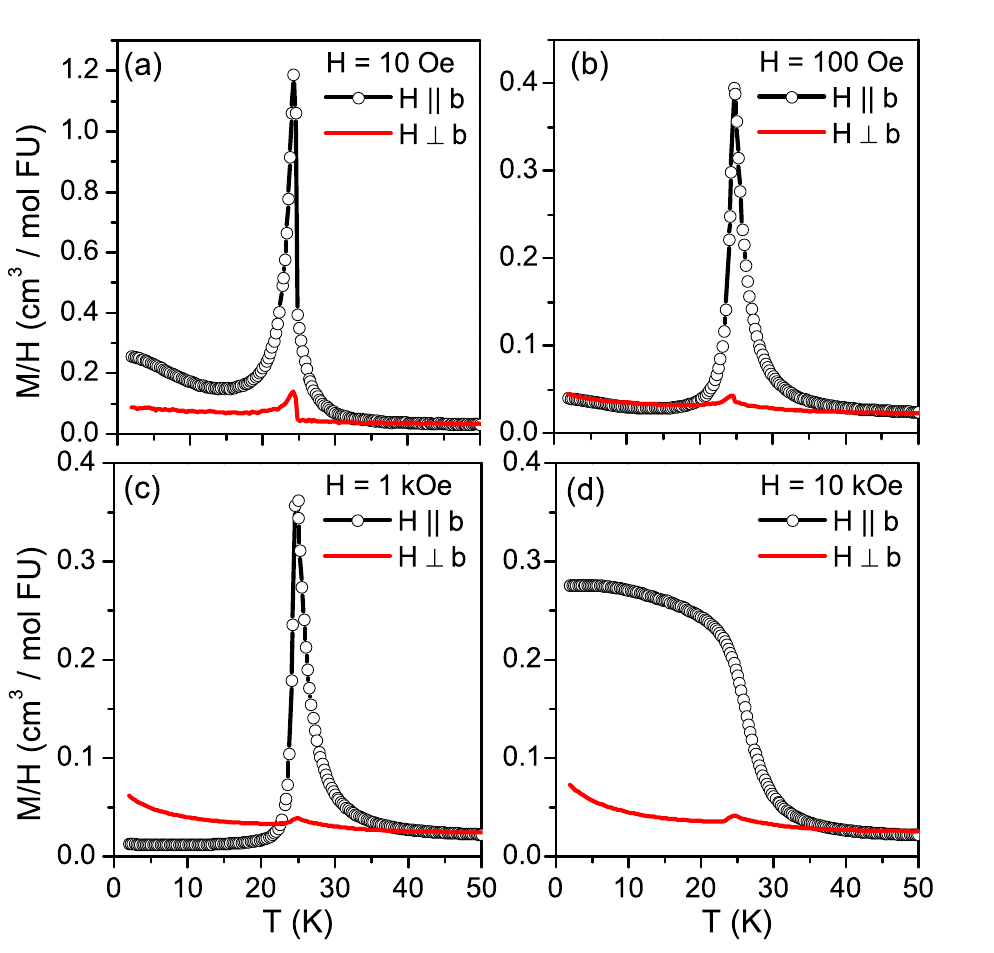}
\caption{\label{fc} (Color online) Magnetization curves for co-align single crystals of Rb$_{2}$Fe$_{2}$O(AsO$_{4}$)$_{2}$ measured on warming at the indicated fields after field-cooling from 50 K. Results are shown for H parallel and perpendicular to the $b$-axis along which the sawtooth chain of Fe are oriented.}
\end{figure}

\begin{figure}[btp]
\includegraphics[width=3.5in]{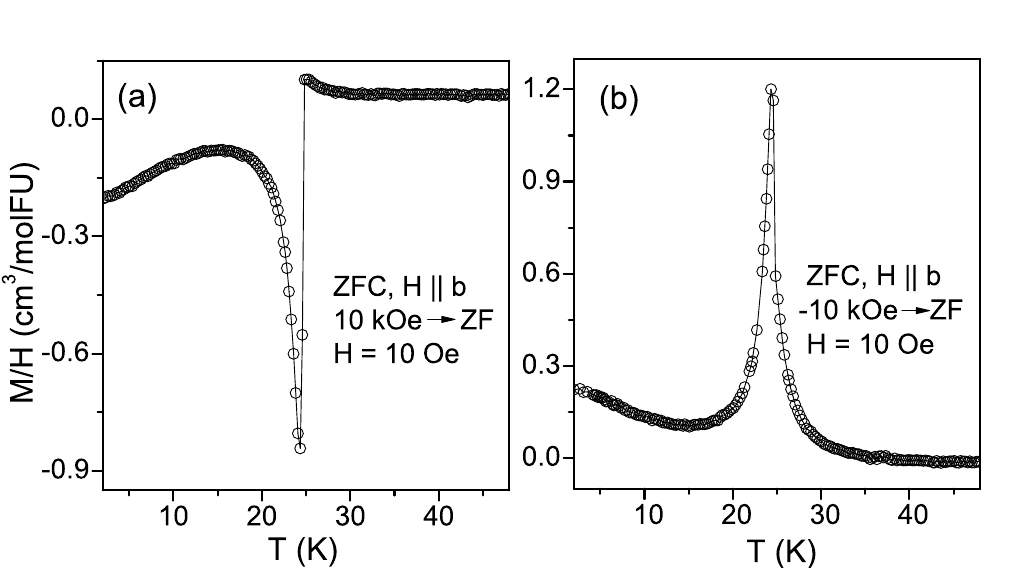}
\caption{\label{zfc} Zero-field-cooled magnetization curves measured in 10 Oe magnetic field (H $\parallel b$) upon warming from 2 to 50 K, after having reduced the field to zero at 50 K from (a) 10 kOe, which leaves a small negative residual field, and (b) -10 kOe which leaves a small positive residual field.}
\end{figure}

Additional magnetization measurements were performed near the ordering temperature. Data were collected upon warming from 2 to 50 K in zero-field-cooled (ZFC) and field-cooled (FC) modes. Figure~\ref{fc} displays the FC results for the four selected fields of 10 Oe, 100 Oe, 1 kOe and 10 kOe. The temperature dependence changes significantly as the applied field is varied, and the data show strong anisotropy between the two measurement directions at all fields. For H along the direction of the sawtooth chains (the $b$-axis), a sharp cusp, suggesting a canted antiferromagnetic order, is seen at the magnetic ordering transition for H $\leq$ 1 kOe, while at 10 kOe the susceptibility increases and then saturates upon cooling, similar to the behavior expected at a ferromagnetic transition. For H aligned perpendicular to the $b$-axis, a smaller cusp is seen at all of the fields shown. The qualitative difference between the behavior in the two directions at H = 10 kOe is due to a field induced magnetic transition which occurs at a different field in the different orientations and is discussed in more detail below.

Negative magnetic susceptibilities were observed in some ZFC measurements at low fields. An example is shown in Figure ~\ref{zfc}(a), where the magnetization is negative at low temperatures, and switches sign abruptly upon warming as the phase transition is approached. This was observed at fields up to 100 Oe and in both orientations of the field. Temperature induced magnetization reversal phenomena is well known to occur in ferrimagnetic systems with strong magnetic anisotropy displaying a partial cancellation of antiferromagnetically coupled magnetic sublattices with different magnitudes of magnetic moments and/or different temperature dependence of magnetization.\cite{Neel,Menyuk} This scenario is certainly plausible in our system where Fe$^{3+}$ occupies two non-equivalent sites and are exposed to different molecular fields. However, a negative ZFC magnetization curve has also been attributed in some cases to the existence of a small trapped field in the superconducting magnet.\cite{Kumar,Belik} Such an experimental artefact is likely to occur in some ferromagnetic materials with strong anisotropy and a significant coercive field that can drastically affect the ZFC process. To investigate this possibility we performed experiments in which we reducing the field to zero from both 10 kOe and -10 kOe using the MPMS "no-overshoot" mode at 50 K, then cooling in "zero field" to 2 K, applied 10 Oe, and then measured upon warming. As discussed in Ref.\onlinecite{Kumar}, the sign of the trapped field is opposite in sign when reducing the field to zero from a positive or negative field, and hence, a starting field of 10 kOe is expected to generate a small negative trapped field, while -10 kOe will create a small positive trapped field. Our measurements on a Pd reference sample using similar conditions confirm this and indicate the magnitude of the field is several Oe. As apparent in Fig.~\ref{zfc}, the ZFC magnetization curves are indeed following the direction of the trapped field in the superconducting magnet. This prompts us to conclude that the observed negative magnetization under ZFC condition is not a ferrimagnetic effect, but rather a feature induced by the presence of pinned uncompensated spins originated from inhomogeneities,\cite{Belik} or from the antiferromagnetic domain walls,\cite{Ueland,Bode} or slight canting in the antiferromagnetic structure, although no canting could be resolved in the neutron diffraction results presented below. We note that several ZFC measurements using a field of 10 Oe were also conducted after resetting or quenching the magnet at 50 K. Some of these measurements resulted in behavior like that shown in Fig.~\ref{zfc}(a), and others like Fig.~\ref{zfc}(b). This indicates that the residual field, which is expected to be very small ($<$ 1 Oe) after resetting the magnet had unpredictable polarity and that the low temperature state that is adopted by the material upon cooling through the transition temperature is extremely sensitive to even very small magnetic fields.

\begin{figure}[tbp]
\includegraphics[width=3.3in]{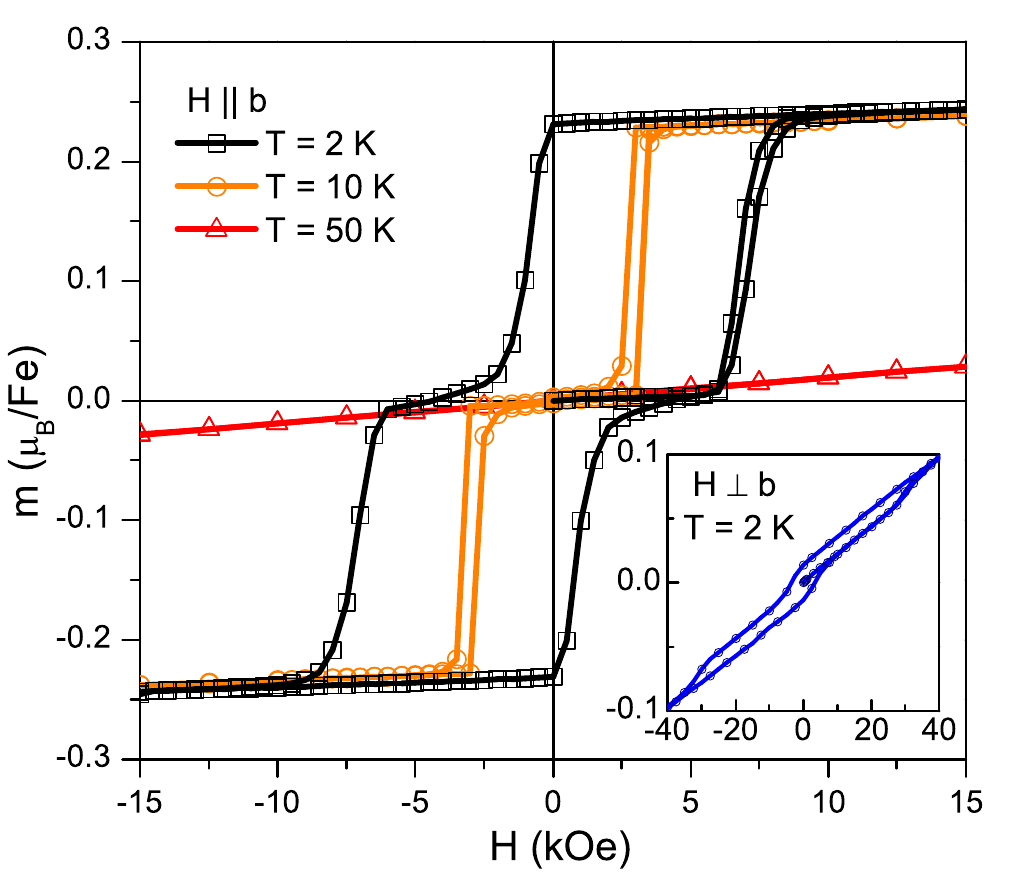}
\caption{\label{hystcrys} (Color online) Magnetic field dependence of magnetization for oriented single
crystals of Rb$_{2}$Fe$_{2}$O(AsO$_{4}$)$_{2}$ with H $\parallel b$ at four different temperatures. Shown in the
insert is the isothermal magnetization curve at 2 K for H $\perp b$ .}
\end{figure}

\begin{figure}[tbp]
\includegraphics[width=3.3in]{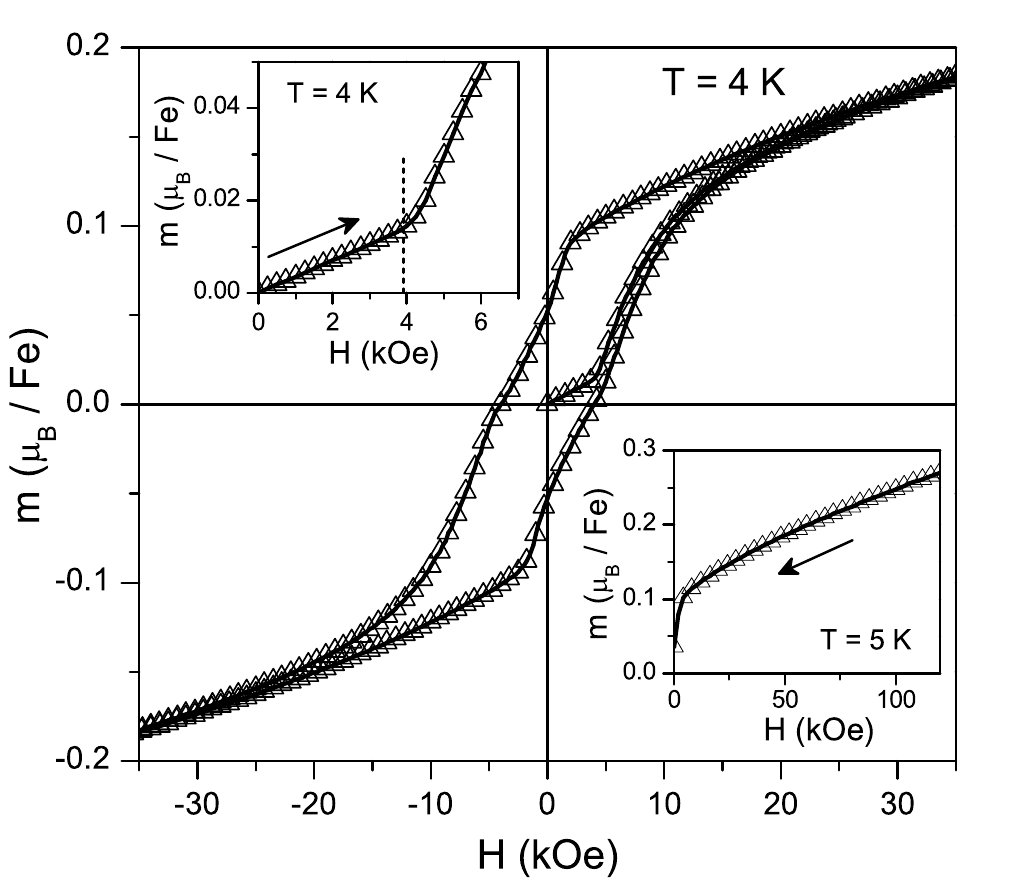}
\caption{\label{hystpow} Isothermal magnetization data measured on a compacted powder sample at T = 4 K.  The upper inset shows a blow up of the region near the metamagnetic transition ( $\approx$4 kOe). The lower insert displays the magnetization data measured upon reducing the field from 120 kOe at 5~K.}
\end{figure}

The isothermal magnetization curves at 2 K, 10 K and 50 K in magnetic fields up to 15 kOe applied along $b$ axis are shown in Fig.~\ref{hystcrys}. The presence of hysteresis in the magnetization data demonstrates the presence of a ferroic component in the high-field magnetic phase. Apart from a hysteresis separating the ascending and descending branches, there is a readily noticeable stepwise magnetic behavior. The metamagnetic transition occurs near 3 kOe at T = 2 K and shifts to 6 kOe at T = 10 K. When the field is applied perpendicular to the $b$ axis the metamagnetic phase transition occurs at a much higher field of 30 kOe, at T = 2 K (as shown in the inset of Fig.~\ref{hystcrys}). As expected, the magnetization measured on compacted powder sample shows a more gradual field transition, arising at approximately 4 kOe at a temperature of 4 K (see Fig.~\ref{hystpow}). It is also worth noting that the magnetization curves does not reach the full saturation even for the highest applied external fields of 120~kOe. The largest absolute value of the magnetization, of about 0.3 $\mu_B$/Fe, is smaller by over an order of magnitude than the predicted value corresponding to a parallel alignment of the Fe$^{3+}$ magnetic moments ($gS$ = 5 $\mu_B$).

\subsection{Neutron diffraction}

\subsubsection{Measurements in zero magnetic field}

\begin{figure}[tbp]
\includegraphics[width=3.4in]{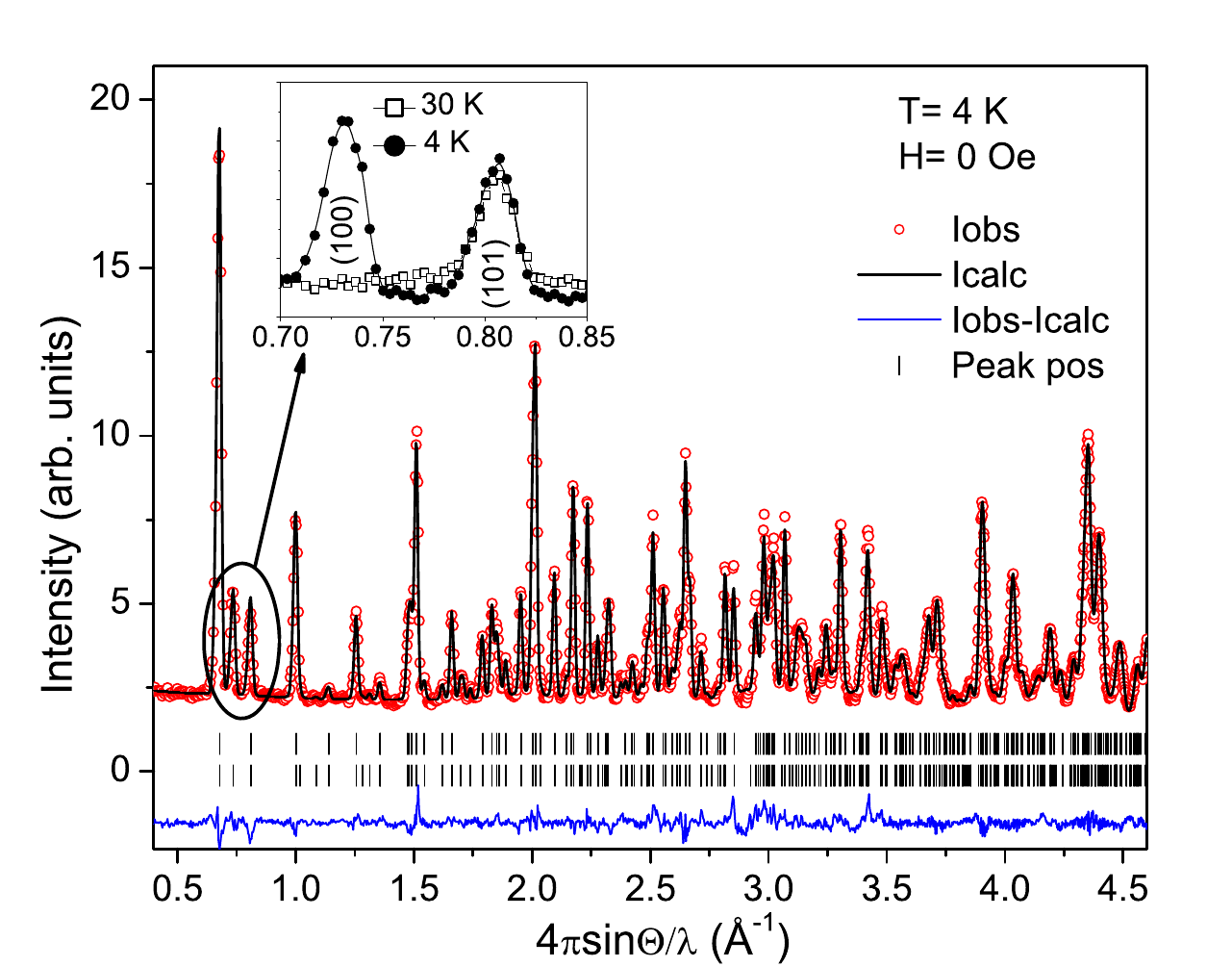}
\includegraphics[width=3.4in]{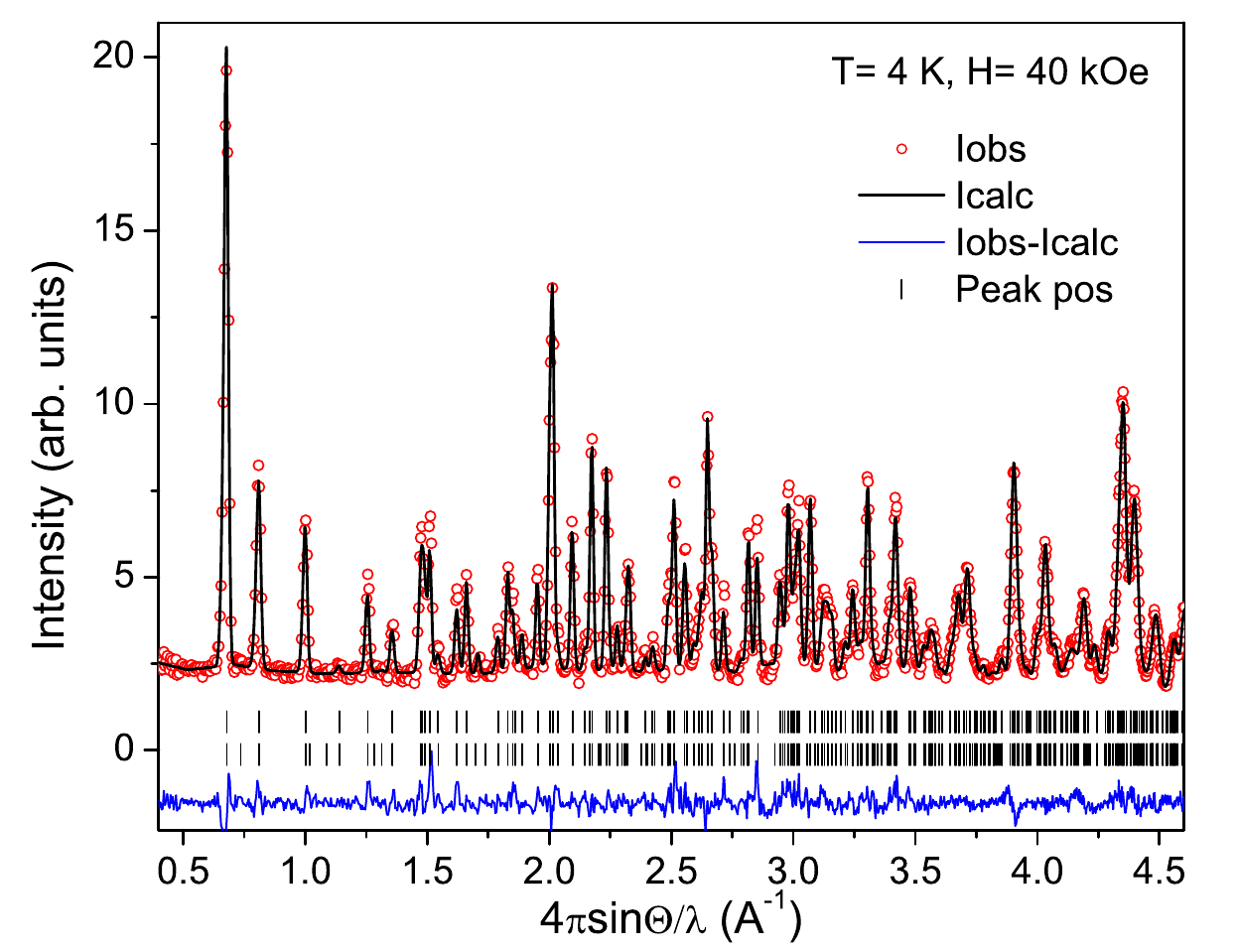}
\caption{\label{rietveld} (Color online) Upper panel: Rietveld plot of neuron powder diffraction data collected at 4 K in zero magnetic field. The inset displays a comparison of the low-\emph{Q} data, near the (100) magnetic peak, measured at 4 K and 30 K. Lower panel: Rietveld plot of neuron data measured at 4 K in magnetic field of 40~kOe.}
\end{figure}

\begin{figure}[btp]
\includegraphics[width=3.5in]{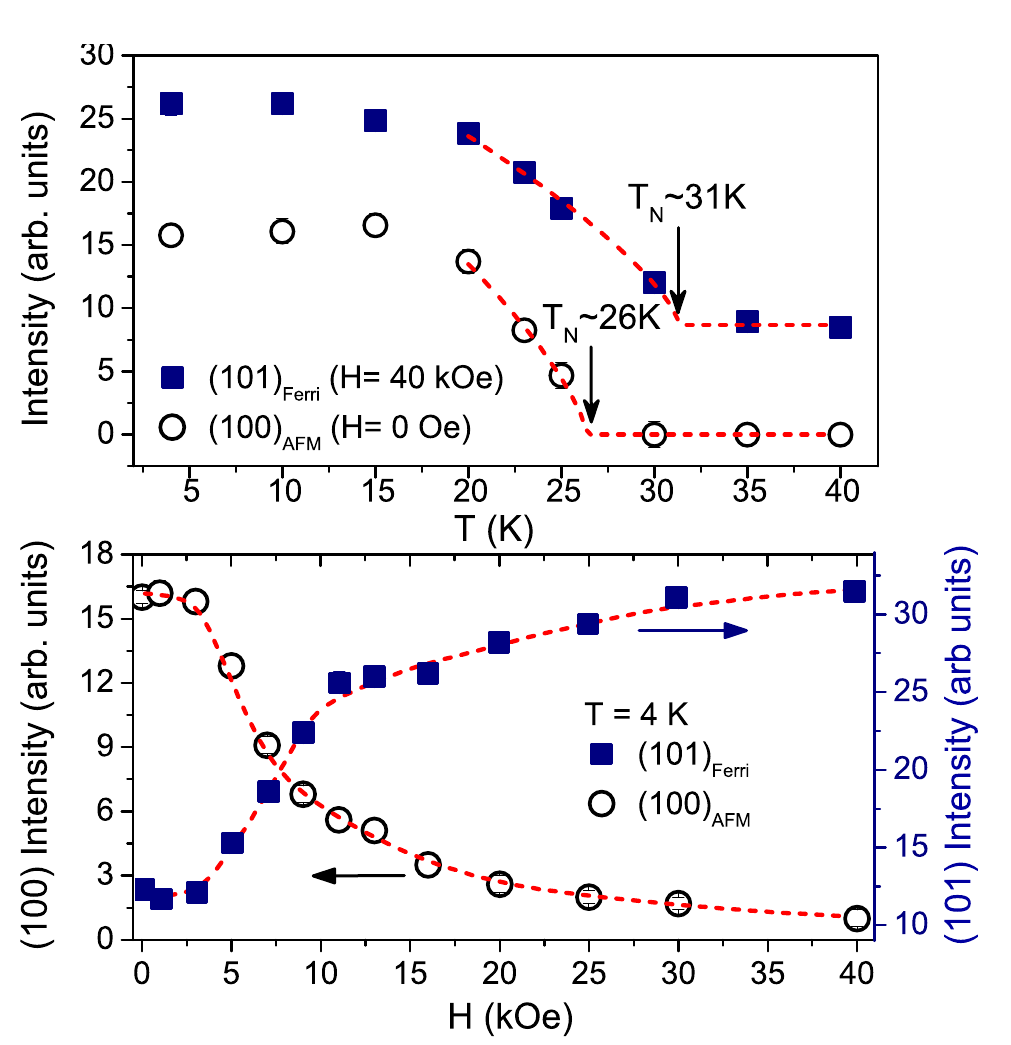}
\caption{\label{orderparam} (Color online) Upper panel: Temperature dependence of the (100) and (101) peaks intensities measured in zero magnetic field and 40 kOe, respectively. The lines represent a guide to the eye indicating power law behavior. Lower panel: Variation with magnetic field of the (100) and (101) magnetic peaks at 4 K. The dashed line is a guide to the eye. The vanishing of (100) reflection at the expense of (101) magnetic is caused by an induced transition from antiferromagnetic to ferrimagnetic state.}
\end{figure}

Neutron diffraction patterns collected in the paramagnetic state at 40 K are well described by the structural
model described in the previous section. Upon cooling below 25 K, magnetic Bragg peaks (e.g. (100), (102)) start to appear at low angles (see Fig.~\ref{rietveld}) indicating a transition to an antiferromagnetic long-rage order. The evolution with temperature of (100) magnetic peak intensity is shown in Fig.~\ref{orderparam}(upper panel). Near the transition temperature the order parameter follows a power law behaviour.

Representational analysis was conducted to determine of the symmetry allowed magnetic structures that can result
from a second-order magnetic phase transition given the crystal structure and the propagation vector $k$=(0,~0,~0) of the magnetic ordering. These calculations were carried out using the program SARA$h$-Representational Analysis.\cite{sarah} There are eight possible irreducible representations (IRs) associated with the $Pnma$ space group and $k$=(0,~0,~0). The decomposition of the magnetic representation for the Fe1 site $( .715,~ .25,~ .565)$ is $\Gamma_{Mag}=1\Gamma_{1}^{1}+2\Gamma_{2}^{1}+2\Gamma_{3}^{1}+1\Gamma_{4}^{1}+1\Gamma_{5}^{1}+2\Gamma_{6}^{1}+2\Gamma_{7}^{1}+1\Gamma_{8}^{1}$,
while for the Fe2 site $( 0,~ 0,~ .5)$ is
$\Gamma_{Mag}=3\Gamma_{1}^{1}+0\Gamma_{2}^{1}+3\Gamma_{3}^{1}+0\Gamma_{4}^{1}+3\Gamma_{5}^{1}+0\Gamma_{6}^{1}+3\Gamma_{7}^{1}+0\Gamma_{8}^{1}$.
The labeling of the IRs follows the scheme used by Kovalev.\cite{Kovalev} Table~\ref{basisvectors} presents only
two of representations ($\Gamma$1, $\Gamma$5) that are relevant for our discussion here.

\begin{table}[tbp]
\caption{\label{basisvectors} Basis vectors (BV) of the IRs  $\Gamma_{1}$ and $\Gamma_{5}$, of the space group
$Pnma$ with $k$ = (0,~0,~0). The equivalent Fe1 atoms of the nonprimitive basis are defined according to 1: $(
.715,~ .25,~ .565)$, 2: $( .215,~ .25,~ .934)$, 3: $( .284,~ .75,~ .434)$, 4: $( .784,~ .75,~ .065)$. The Fe2 atoms are defined as follows: 1: $( 0,~ 0,~ .5)$, 2: $( .5,~ .5,~ 0)$, 3: $( 0,~ .5,~ .5)$, 4: $( .5,~ 0,~ 0)$.}
\begin{ruledtabular}
\begin{tabular}{ccc|ccc|ccc}
  IR  &  BV  &  Atom & \multicolumn{3}{c}{Fe1}& \multicolumn{3}{c}{Fe2}\\
      &      &             &$m_{\|a}$ & $m_{\|b}$ & $m_{\|c}$ &$m_{\|a}$ & $m_{\|b}$ & $m_{\|c}$ \\[3pt]
\hline
$\Gamma_{1}$ & $\bfpsi_{1}$ &      1 &      0 &      1 &      0 &      0 &      1 &      0  \\[3pt]
             &              &      2 &      0 &     -1 &      0 &      0 &     -1 &      0  \\[3pt]
             &              &      3 &      0 &      1 &      0 &      0 &      1 &      0  \\[3pt]
             &              &      4 &      0 &     -1 &      0 &      0 &     -1 &      0  \\[3pt]
             & $\bfpsi_{2}$ &      1 &        &        &        &      1 &      0 &      0  \\[3pt]
             &              &      2 &        &        &        &      1 &      0 &      0  \\[3pt]
             &              &      3 &        &        &        &     -1 &      0 &      0  \\[3pt]
             &              &      4 &        &        &        &     -1 &      0 &      0  \\[3pt]
             & $\bfpsi_{3}$ &      1 &        &        &        &      0 &      0 &      1  \\[3pt]
             &              &      2 &        &        &        &      0 &      0 &     -1  \\[3pt]
             &              &      3 &        &        &        &      0 &      0 &     -1  \\[3pt]
             &              &      4 &        &        &        &      0 &      0 &      1  \\[3pt]
\hline
             &              &        &        &        &        &        &        &         \\
$\Gamma_{5}$ & $\bfpsi_{4}$ &      1 &      0 &      1 &      0 &      0 &      1 &      0  \\[3pt]
             &              &      2 &      0 &      1 &      0 &      0 &      1 &      0  \\[3pt]
             &              &      3 &      0 &      1 &      0 &      0 &      1 &      0  \\[3pt]
             &              &      4 &      0 &      1 &      0 &      0 &      1 &      0  \\[3pt]
             & $\bfpsi_{5}$ &      1 &        &        &        &      1 &      0 &      0  \\[3pt]
             &              &      2 &        &        &        &     -1 &      0 &      0  \\[3pt]
             &              &      3 &        &        &        &     -1 &      0 &      0  \\[3pt]
             &              &      4 &        &        &        &      1 &      0 &      0  \\[3pt]
             & $\bfpsi_{6}$ &      1 &        &        &        &      0 &      0 &      1  \\[3pt]
             &              &      2 &        &        &        &      0 &      0 &      1  \\[3pt]
             &              &      3 &        &        &        &      0 &      0 &     -1  \\[3pt]
             &              &      4 &        &        &        &      0 &      0 &     -1  \\[3pt]
\end{tabular}
\end{ruledtabular}
\end{table}

\begin{figure}[btp]
\centering
\subfigure[]{\includegraphics[width=3.0in]{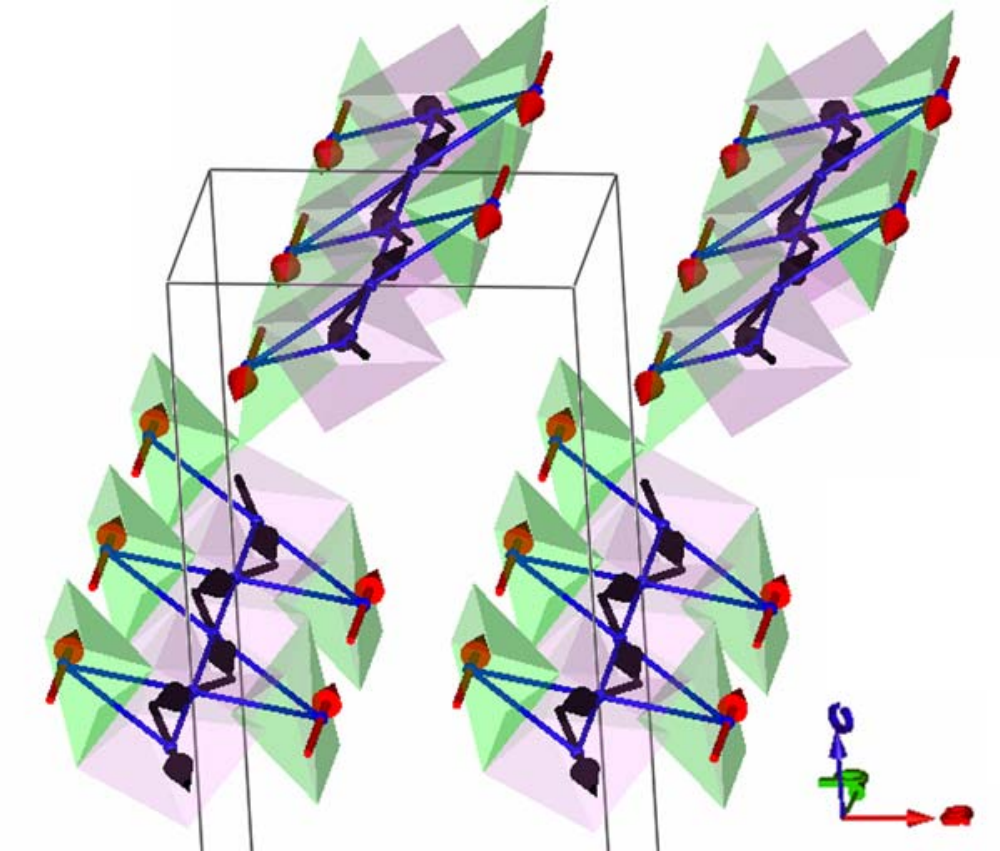}}
\subfigure[]{\includegraphics[width=3.0in]{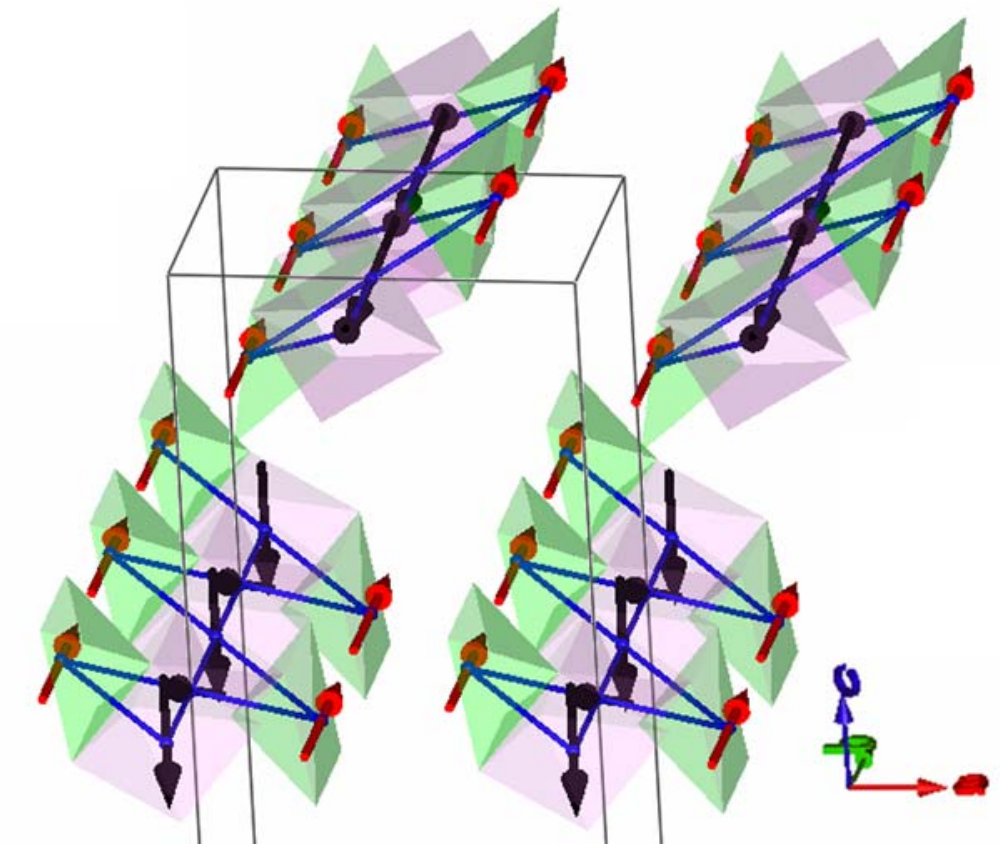}}
\caption{\label{magstruct} (Color online) (a) Schematic view of the magnetic structure of Rb$_{2}$Fe$_{2}$O(AsO$_{4}$)$_{2}$ at 4 K and zero magnetic field. The Fe1 magnetic moments are collinear with the $b$-axis, while the Fe2 moments are forming undulating patterns, running in opposite direction with respect to the Fe1 moments. The ferrimagnetic chains are coupled antiferromagnetically to neighbouring chains in the $c$ direction (b) Magnetic structure at 4 K and applied magnetic filed of 40 kOe. In contrast to the zero-field magnetic order, the polarity of the ferrimagnetic chains remains the same throughout the lattice yielding an overall ferrimagnetic ordering.}
\end{figure}

The 4 K diffraction data was fit by a magnetic structure model based on the representation $\Gamma_{1}$, defined in Table~\ref{basisvectors}. $\Gamma_{1}$ representation is equivalent to the Shubnikov magnetic space group $Pnma$. Within this model, the Fe1 moments are constrained to lie along the $b$-axis direction and are ferromagnetically coupled to each other inside the chain. In contrast, the Fe2 moments have components in all three coordinate directions, and form an undulating pattern with the relative angles between neighboring spins of approximately 60$^\circ$. The spin canting occurs in such a way that they remain mostly confined to the plane of the sawtooth chain. The projection along the chain axis of the Fe2 moment is oriented in opposite direction to the Fe1 moment, thus producing a non-zero net moment. The polarity of the chains alternates along the $c$-axis direction, giving an overall antiferromagnetic arrangement. A stereographic view of the magnetic structure is illustrated in Fig.~\ref{magstruct}(a).

The refined magnitudes of the ordered moments for Fe1 and Fe2 sites are 3.64(6) and 3.19(8)~$\mu_B$, respectively. The projections of Fe2 magnetic moment on the three axes of coordinates are: $\mu_a$ = 1.19(8)~$\mu_B$, $\mu_b$ = 2.88(4)~$\mu_B$ and $\mu_c$ = 0.68 (9)~$\mu_B$. The refined values are significantly lower than those expected for the Fe$^{3+}$ ions with the effective spin $S$=5/2. Considering the high value of the frustration parameter $f = |\Theta_{CW}|/T_N$ of approximately 18, one can infer that the reduced static magnetic moment is due to the presence of spin frustration. The frustration is reasonably expected in this system due to competing superexchange interactions within the triangular chains. Furthermore, the non-collinear spin arrangement is reminiscent of the predicted non-collinear spin configurations in triangular-lattice Heisenberg antiferromagnet where the frustration of the three nearest-neighbor spins on a triangular plaquette is resolved by a 120$^\circ$ rotation of neighboring spins.\cite{Capriotti}

Perhaps the most important aspect that needs to be emphasized here is the fully
compensated character of the magnetic structure, which makes the magnetization reversal process observed in the ZFC magnetization data difficult to comprehend. Ferrimagnetic-like response has previously been observed only in a handful of antiferromagnetic compounds where a weak ferromagnetic moment is due to the tilting of sublattice
magnetizations.\cite{Kageyama,Singh,Ren,Kimishima} One of the mechanisms that has been proposed to explain the
magnetization reversal in antiferromagnets relies on the competition between single ion anisotropy and the
Dzyaloshinsky–Moriya (D-M) interaction. For our system, the antisymmetric D–M exchange coupling can exist between the magnetically ordered Fe2 ions located at $4b$ site of $Pnma$, and the magnetic spin corresponding to that site has indeed been determined to be canted. However, because the chains are antiferromagnetically coupled, any ferromagnetic component due to canting is compensated by the nearest neighbor chain. These results come to confirm our conjecture in the previous section that the negative magnetization is not an inherent ferrimagnetic feature, but likely an effect of an uncompensated magnetization originating from antiferromagnetic domain walls or from other magnetic inhomogeneities present in the sample.

\subsubsection{Measurements in external magnetic fields}

To understand the effect of magnetic field on the magnetic structure of Rb$_{2}$Fe$_{2}$O(AsO$_{4}$)$_{2}$,
polycrystalline sample in the form of pressed pellets has been measured under different  magnetic fields up to 40~kOe. The diffraction data measured at 4~K reveals a progressive decrease in the amplitude of (100) magnetic peak with increasing the applied magnetic field, and at the same time, an increase in intensity of the (101) peak. The change in peaks intensities only manifests at low momentum transfer ($Q$), in accordance with the Fe$^{3+}$ magnetic form factor. The fact that the high-$Q$ reflections are not affected by the magnetic field gives assurance that the sample grains do not re-align in the direction of the field. As displayed in Fig.~\ref{orderparam}(lower panel), the (100) peak intensity starts to decrease just above 3~kOe, and it nearly vanished at 40~kOe. The (101) peak follows a very similar trend to that seen for the isothermal magnetization curve measured for the powder sample in the magnetically ordered state at 4 K, in Fig.~\ref{hystpow}. The intensity of the (101) peak has also been monitored as a function of temperature in a constant magnetic field of 40~kOe. The order parameter versus temperature shown in Fig.~\ref{orderparam}(upper panel) follows a power-law dependence, similar as that observed for the (100) peak in zero magnetic field.

Rietveld refinement has been performed using the diffraction data collected at 40 kOe and 4 K, and the fit result is presented in Fig.~\ref{rietveld}. The magnetic scattering has been well reproduced by a ferrimagnetic model based on the irreducible representation $\Gamma_5$ given in Table~\ref{basisvectors}. This representation is equivalent to the Shubnikov magnetic space group $Pn^\prime ma^\prime$. In this new magnetic state all chains exhibit the same polarity, with all symmetry equivalent Fe ions having the $b$-axis component of magnetic moment aligned in the same direction. On the other hand, Fe1 and Fe2 sub-lattices maintain opposite magnetizations while not compensating each other. Another effect of the magnetic field is the change in the canting direction of Fe2 spins with respect to the local plane of the sawtooth-chain. The spins rotate around the $b$-axis from an in-plane canting to an out-of-plane canting. The relative angle between Fe2 and Fe1 moments remains close to 120$^\circ$. The high-field magnetic structure is shown in Fig.~\ref{magstruct}(b).

The magnitude of the ordered magnetic moments at 40~kOe is determined to be very close to that observed for the
antiferromagnetic state in zero magnetic field. The refined Fe1 moment is 3.5(1)~$\mu_B$, while the amplitude of the Fe2 moment is 3.1(1)~$\mu_B$ with the three axes components $\mu_a$ =0.5(1)~$\mu_B$, $\mu_b$ = 2.76(6)~$\mu_B$ and $\mu_c$ = 1.3(1)~$\mu_B$. These values appear to rule out the possibility
that the external magnetic field may polarize additional disordered Fe spins into the novel ordered state. On the other hand, the net magnetization resulted from such amplitudes is about 0.37~$\mu_B$ per Fe$^{3+}$, which is reasonably consistent to what has been determined from magnetization measurements (see Figs.~\ref{hystcrys} and \ref{hystpow}). Interestingly, the magnetization data shown in Fig.~\ref{hystcrys} suggest that at sufficiently low temperatures, the field induced ferrimagnetic state can be maintained even after reducing the field back to zero.

It appears that the metamagnetic transition corresponds to the transition from the antiferromagnetic to ferrimagnetic state where all magnetic couplings between adjacent sawtooth chain become exclusively ferromagnetic while the intra-chain exchange interactions remain nearly unaltered. The field induced ferrimagnetic ordering provides an interesting insight into the tendency of the system to produce a net ferromagnetic component, and opens up the question whether local ferrimagnetic cluster may start to form within the antiferromagnetic host even at very low magnetic fields. The contribution to the scattering of such weak uncompensated magnetic order could be well below the detection limit of our neutron scattering measurements. Magnetic clustering could plausibly explain the observed negative magnetization but it implies the existence of inhomogeneities in the sample. Note that in our case the inhomogeneities would likely have their origin in oxygen vacancies, which could not be either confirmed or ruled out by our structural refinement. As mentioned before, the uncompensated spins might also result from antiferromagnetic domain walls, as demonstrated in the case of the HoMnO$_{3}$ multiferroic system by means of polarized small angle neutron scattering experiments.~\cite{Ueland}

\section{Summary}

The magnetic properties of the new oxy-arsenate Rb$_{2}$Fe$_{2}$O(AsO$_{4}$)$_{2}$ have been studied by means of
magnetization and neutron powder diffraction measurements. The system structure consists of triangle-based magnetic chains made of $S$= 5/2 Fe$^{3+}$ ions occupying two crystallographically distinct tetrahedral sites (Fe1 and Fe2). Magnetization measurements indicate a complex magnetic behavior with a temperature induced magnetization reversal at low magnetic fields, and a step-like metamagnetic transition. Neutron diffraction experiments performed in zero magnetic field reveal an antiferromagnetic long-range magnetic order below the N\'{e}el temperature of about 25~K. The magnetic structure consists of ferrimagnetic sawtooth chains which are mutually compensating each-other along the $c$-direction. Within each chain, the Fe1 moments are collinear lying along the $b$ direction, while the Fe2 moments are reversely canted by approximately 30$^\circ$ forming a zigzag pattern. A reduced static magnetic moments of Fe has been attributed to a magnetic frustration caused by the triangle geometry. Neutron data collected under externally applied magnetic field reveal a transition from an antiferromagnetic to a ferrimagnetic state. In the ferrimagnetic state the individual chains remain ferrimagnetic with similar spin topology but their coupling becomes exclusively ferromagnetic. The ferrimagnetic ordering at  at magnetic fields greater than 3 kOe cannot be invoked to explain the negative ZFC magnetization observed in the macroscopic study. Instead, the present data suggests that the negative magnetization could be an effect of a phase-segregated state with ferrimagnetic inclusions, or of an uncompensated magnetization arising within antiferromagnetic domain walls. On the basis of our results, the sawtooth-like lattice model displayed by the Rb$_{2}$Fe$_{2}$O(AsO$_{4}$)$_{2}$ system deserves further experimental and theoretical investigations to fully understand its magnetic behaviour. While the magnetic phase segregation antiferromagnetic domain walls are at present only hypothetical scenarios, future polarized neutron scattering experiments able to provide increased sensitivity to week ferromagnetism may shed light on the origin of the negative magnetization phenomenon.

\begin{acknowledgments}
Work at the Oak Ridge National Laboratory, was sponsored by the Scientific User Facilities Division (neutron diffraction) and Materials Sciences and Engineering Division (magnetization measurements), Office of Basic Energy Sciences, US Department of Energy (DOE). The authors acknowledge the financial support by grants from the National Science Foundation: DMR-0706426, CHE-9808165 and CHE-9808044. L.D.S. and D.S. travel to ORNL was supported by DOE through the EPSCoR Grant, DE-FG02-08ER46528.
\end{acknowledgments}

\end{document}